# A Model of the Enhanced Vertical Mixing Induced by Wind-Waves

## V. G. Polnikov[1]

[1]A.M. Obukhov Institute of Atmospheric Physics of RAS, Moscow, 119017 Russia;
Corresponding author: Vladislav Polnikov (polnikov@mail.ru)

**Key Points:**

- In the Navier-Stokes equations, a current is decomposed into four constituents, what allows separating the background and wave-induced Reynolds stresses.

- The explicit form of the wave-induced mixing function is found by the Reynolds-stress closure with using the turbulent viscosity in the wave-zone of the interface.

- The mixing function is linear in wave amplitude, resulting in enhanced impact of waves on a vertical mixing and related global geophysical processes.

**Abstract**

In the Navier-Stokes equations, a current is decomposed into four constituents: the mean flow, wave-orbital motion, wave-induced-turbulent and background-turbulent currents. Under certain statistical assumptions, this allows to separate the wave-induced Reynolds stress, $R_w$, from the background one, $R_b$. To close $R_w$, the Prandtl approach for the wave-induced fluctuations is used, resulting in the implicit expression for the wave-induced vertical mixing function, $B_v$. Expression for $B_v$ is specified, basing on the author's results for the turbulent viscosity, found earlier in the frame of the three-layer concept for the air-sea interface. The explicit expression for function $B_v(a, u_*, z)$ is linear in both wave amplitude $a(z)$ at depth $z$ and friction velocity $u_*$ in the air. Due to depth-dependence $a(z) \sim exp(kz)$, the found result for $B_v(a)$ means the possibility of the enhanced impact of waves on the vertical mixing and related geophysical processes, compared with the known cubic wave-amplitude dependence $B_v(a)$.

# 1 Introduction

The adequate description of the upper-ocean mixing processes is very important for the ocean-circulation modelling from both scientific and applied points of view (Kitaigorodskii & Lamley, 1983; Kantha & Clayson, 1994; Ezer, 2000; Mellor, 2001; Qiao et al., 2004, 2010). The scientific interest to the problem is determined by the natural aspiration of researchers to clarify the physics of mixing processes in the upper ocean. The practical significance of the problem solution is stipulated by the tasks of improving both the ocean-circulation simulations and forecasting the weather and climate variability. The certain geophysical applications of such solutions are well represented in the recent papers (Qiao et al., 2016, Aijaz et al., 2017; Walsh et al., 2017).

Here we mention only that the traditional approach to the problem is based on the using the multi-level schemes of the statistical moments closure, following to the formulations systemized by Mellor and Yamada (1982). In this approach, they consider directly the full Navier-Stokes equations containing motions of different scales of variability. Accounting for details of this variability leads to the necessity to consider equations for statistical moments, describing different turbulent variables, what is typical for the theory of turbulence (Monin & Yaglom, 1971). On this way, an evident progress was achieved (e.g., Anis & Moum, 1995; Ardhuin & Jenkins, 2006; Janssen, 2012). Herewith, the Mellow-Yamada approach contains numerous physical assumptions, hypotheses and simplifications. Eventually, all these uncertainties result in the underestimation of the vertical mixing in the upper layer and the mixed layer depth (Kantha and Clayson, 1994; Ezer, 2000; Mellor, 2000, Qiao et al., 2010, Janssen, 2012, among others).

These defects of the ocean-circulation models could be reasonably corrected by means of involving into consideration the surface-wave influence on vertical mixing processes. This idea was realized in a lot of works (Kitaigorodskii & Lamley, 1983; Craig & Banner, 1994; Anis & Moum, 1995; Ezer, 2000; Mellor, 2003; Ardhuin & Jenkins, 2006; Janssen, 2012, and references therein), where the authors tried to use different forms of involving the processes at the air-sea interface into the traditional turbulence schemes. All the authors stated that surface waves can enhance the mixing in the upper ocean (e.g., Qiao et al., 2016, Aijaz et al., 2017; Walsh et al., 2017). However, the task of effective parameterization processes accompanying the wave-induced mixing in the three-dimensional circulation models is remaining in progress.

In this regard, one of the successful results of solving the problem of the upper-ocean mixing was achieved in (Qiao et al., 2004), where they have analyzed the wave-induced turbulence by the direct closing the Reynolds stresses, what formally corresponds to the first-level closure scheme in the theory of turbulence (Monin & Yaglom, 1971). This approach is based on the concept of existence of the wave-induced turbulence as the addition to the background turbulence in the upper ocean. This concept was justified and confirmed in numerous experimental and theoretical studies (Gemmrich & Farmer, 2004; Gemmrich, 2010; Babanin 2006; Babanin and Haus, 2009; Dai et al., 2010; Pleskachevsky et al., 2011; Babanin & Chalikov, 2012; Benilov, 2012; Sutherland and Melville, 2015, and references therein). Among these works, the analytical approach applied by Qiao et al. (2004) is seemed as the simplest, direct, and rather effective solving the task under consideration. Therefore, here we dwell on it in some details.

Qiao et al. (2004) have started their analysis of the wave-induced mixing directly from the closure of the Reynolds stresses. Decomposing current fluctuation into the wave-induces one, $\tilde{u}_i^{'}$, and turbulent fluctuation existing without waves, $u_i^{'}$, they concentrated on stress $<\tilde{u}_i^{'}u_j^{'}>$ (brackets $<...>$ mean the statistical averaging) in the simplest case of the uniform mean current, $\overline{\mathbf{U}} = (U,0,0)$. For the vertical mixing stress, $<u_1^{'}\tilde{u}_3^{'}>$, they assumed the following closure

$$<u_1^{'}\tilde{u}_3^{'}> = B_v \frac{dU}{dz} \qquad . \tag{1}$$

Here sub-indexes (1,2,3) denote x-, y-, z- components of currents, and $B_v$ is defined as the wave-induced vertical turbulent viscosity (vertical mixing function). In their work, $B_v$ was assumed as

$$B_v = <\tilde{\lambda}_3^{'}\tilde{u}_3^{'}>, \tag{2a}$$

where $\tilde{\lambda}_3^{'}$ is the Prandtl mixing length of the wave-induced turbulence. After that, the wave-induced fluctuation, $\tilde{u}_3^{'}$, was expressed via the wave-orbital current, $\tilde{u}_1^{'}$, by means of the Prandtl approximation

$$\tilde{u}_3^{'} = \tilde{\lambda}_3^{'}( d\tilde{u}_1^{'} / dz ) \qquad . \tag{2b}$$

Then, basing on the potential theory for surface waves, values $\tilde{\lambda}_3^{'}$ and $\tilde{u}_1^{'}$ in Eqs. (2a,b) were expressed via the wave-number spectrum of waves, $S(\mathbf{k})$, and the final expression for the wave-induced mixing function got the form (see details in the original)

$$B_v(S,z) = \alpha \int S(\mathbf{k}) \exp(2kz) d\mathbf{k} \cdot \frac{d\left(\int \omega^2(\mathbf{k}) S(\mathbf{k}) \exp(2kz) d\mathbf{k}\right)^{1/2}}{dz} . \tag{3}$$

In (3), $\alpha$ is the dimensionless fitting coefficient, $\omega(\mathbf{k})$ is the wave frequency corresponding to wave vector $\mathbf{k}$ for gravity waves, and the exponents denote the depth dependence of spectrum.

The main feature of Eq. (3) is the cubic dependence of function $B_v(S,z)$ on the wave amplitude on the surface, $a_0 = \left(\int S(\mathbf{k}) d\mathbf{k}\right)^{1/2}$, resulting in the rather strong depth-decay: $B_v(S,z) \propto \exp(3k_p z)$ ($k_p$ is the wave number corresponding to the peak of spectrum). Despite of this strong dependence on depth, "*adding $B_v$ to the vertical diffusivity in a global ocean-circulation model yields a temperature structure in the upper 100 m that is closer to the observed climatology than in a model without the wave-induced mixing*" (the citation from Qiao et al.,

2004). Later, the fact of remarkable impact of the wave-induced mixing on the global circulation was confirmed in the series of works (e.g., Dai et al., 2010; Qiao et al., 2010, Pleskachevsky et al., 2011; Huang et al., 2012; Qiao et al., 2016; Aijaz et al., 2017; Walsh et al., 2017, and references therein).

According to the said, the presented approach is seemed as a very prospective semi-phenomenological solution of the wave-induced mixing problem. Though, taking into account several vulnerable assumptions in the presented version (Eqs. 2a,b), this approach is worthwhile to be elaborated with the aim of improving some details.

The present paper is aimed to constructing a new version of the approach. Eventually, it turned out that the new version for the wave-induced turbulence results in a reasonably greater impact of waves on the vertical mixing then it is described by Eq. (3). It means that surface wind waves can make a much more impact on all geophysical processes related to a vertical mixing in oceans than it is known already from the referenced papers (i.e., Qiao et al., 2016; Aijaz et al., 2017; Walsh et al., 2017).

## 2 Main derivations

### 2.1 Theoretical backgrounds

With the aim of representing some details in the convincing form, here we reproduce some known analytical derivations introducing into the problem under consideration. Herewith, one should keep in mind that we deal with the motions in the upper-water layer, located deep enough below the wavy surface, where the background currents exist independently of the surface waves.

Following to the well-known approach of the theory of turbulence (Monin & Yaglom, 1971; Phillips, 1977; see also Anis & Moum, 1995), we start from the Navier-Stokes equations written in the tensor form

$$\frac{\partial U_i}{\partial t} + U_j \frac{\partial U_i}{\partial x_j} = -\frac{1}{\rho}\frac{\partial P}{\partial x_i} + \nu \frac{\partial^2 U_i}{\partial x_j^2} \quad . \tag{4}$$

Here $U_i$ ($i$=1,2,3) is the x-y-z-component of the current in the upper layer below the deepest wave troughs, $P$ is the pressure, $\rho$ is the water density, and $\nu$ is the kinematic viscosity of the water. (Repeating indexes mean the summation). Following to (Anis & Moum, 1995; Qiao et al., 2004), the Coriolis term is not included in (4) for the reason of the small-scale motions considered in this task.

Before making statistical averaging, it is very important to separate accurately different kinds of motions. To this end, we put the following decomposition for the current

$$U_i = \bar{U}_i + u_i = \bar{U}_i + \tilde{u}_i + \tilde{u}_i' + u_i' \tag{5}$$

Here $\bar{U}_i$ is the mean current without waves, $u_i$ is the total addition to the mean current, $\tilde{u}_i$ is the wave orbital current, $\tilde{u}_i'$ is the wave-induced turbulent current, and $u_i'$ is the turbulent current without waves (later, $u_i'$ is called as the "background" turbulence), Note that the background turbulent motions are independent of the turbulent motions induced by waves, though, both fluctuations may correlate statistically each with other. This statement is very important for the following consideration.

The same decomposition is assumed for pressure $P$. But below, we shall not touch the pressure-terms in Eq. (4), supposing that, in this version of our constructions, they do not directly effect on the wave-induced vertical mixing under consideration.

Now, the following statistical approximations are accepted:

1) All the wave- and turbulent-current summands in (5) vanish in the mean:

$$< u > = 0;$$ (6a)

2) There is no correlation between the mean, wave, and turbulent currents:

$$< \overline{U}_i \tilde{u}_j > = 0, \ < \overline{U}_i \tilde{u}_j^{'} > = 0, \ < \overline{U}_i u_j^{'} > = 0, \ < \tilde{u}_i u_j^{'} > = 0, \ < \tilde{u}_i u_j^{'} > = 0.$$ (6b)

3) The wave-induced-turbulent and background-turbulent summands may correlate:

$$< \tilde{u}_i^{'} \tilde{u}_j^{'} > \neq 0, \ < \tilde{u}_i^{'} u_j^{'} > \neq 0, \ \text{and} \ < u_i^{'} u_j^{'} > \neq 0,$$ (6c)

Additionally, the condition of continuity takes place for each constituent in (5): $\partial U_i / \partial x_i = 0$.

Putting (5) into (4), making ensemble averaging over hundreds of wave periods (denoted as $< \ldots >$), and taking into account ratios (6a,b), one gets the Reynolds equations

$$\frac{\partial \overline{U}_i}{\partial t} + \overline{U}_j \frac{\partial \overline{U}_i}{\partial x_j} + \frac{\partial < u_j u_i >}{\partial x_j} = \left[ pressure - terms \right] + \nu \frac{\partial^2 \overline{U}_i}{\partial x_j^2} \ ,$$ (7)

where the third term in the l.h.s. contains the Reynolds stress $< u_j u_i >$. In view of the above decomposition, the full expression for the stress is the following

$$< u_j u_i > = < \tilde{u}_j \tilde{u}_i > + < \tilde{u}_j \tilde{u}_i^{'} > + < \tilde{u}_j u_i^{'} > +$$
$$+ < \tilde{u}_j^{'} \tilde{u}_i > + < \tilde{u}_j^{'} \tilde{u}_i^{'} > + < \tilde{u}_j^{'} u_i^{'} > + < u_j^{'} \tilde{u}_i > + < u_j^{'} \tilde{u}_i^{'} > + < u_j^{'} u_i^{'} > \ .$$ (8)

Keeping in mind ratios (6 b,c), from (8) one may get

$$< u_j u_i > = < \tilde{u}_j \tilde{u}_i > + < \tilde{u}_j^{'} \tilde{u}_i^{'} > + < \tilde{u}_j^{'} u_i^{'} > + < u_j^{'} \tilde{u}_i^{'} > + < u_j^{'} u_i^{'} > \ .$$ (9)

If there are no waves, i.e. when $\tilde{u}_i = 0$, and $\tilde{u}_i^{'} = 0$, one has the standard Reynolds stress

$$< u_j u_i > = < u_j^{'} u_i^{'} > \ ,$$ (10)

corresponding to the background turbulence existing at the absence of waves.

In the approximation of the first-level closure (Monin & Yaglom, 1971), from (10) it follows

$$< u_j u_i > = K_{ji} \partial \overline{U}_i / \partial x_i,$$ (11)

where coefficient $K_{ji}$ has the meaning of the turbulent viscosity (at the depth considered). For constant $K_{ji}$, closure (11) leads to the standard form of the viscous term in Eq. (4):

$$\frac{\partial < u_j^{'} u_i^{'} >}{\partial x_j} := \frac{\partial}{\partial x_j} ( K_{ji} \frac{\partial \overline{U}_i}{\partial x_i} ) = K_{ji} \frac{\partial^2 \overline{U}_i}{\partial x_j \partial x_i} \ .$$ (12)

By this way, making the closure of the wave-induced terms in full stress (9), one may get the wave-induced part of the turbulent viscosity resulting in the proper vertical mixing function.

### 2.2 Initial specifications

For simplicity, let us consider the case of uniform and homogeneous current $\mathbf{U}$ with the vertical shear, directed along the OX-axis, i.e.: $\overline{\mathbf{U}} = ( \overline{U}(z) , 0, 0 )$, and $\partial \overline{U} / \partial x_i = 0$ for $i = 1, 2$. Putting $j = 1$ and $i = 3$ in Eq. (9), one gets the stress under consideration of the form:

$$< u_1 u_3 > = < \tilde{u}_1 \tilde{u}_3 > + < \tilde{u}_1^{'} \tilde{u}_3^{'} > + < \tilde{u}_1^{'} u_3^{'} > + < u_1^{'} \tilde{u}_3^{'} > + < u_1^{'} u_3^{'} > ,$$ (13)

which is to be specified.

First, we assume that the wave-orbital motion $\tilde{u}_i$ is potential. Thus, in the two-dimensional ($x$, $z$)-space, a monochromatic wave with amplitude $a$, frequency $\omega$, and wave number $k$, propagating in the deep water, can be described in the linear approximation by the following ratios (Yuan et al., 2013):

the surface elevation is given by

$$\eta(x,t) = a\cos(kx - \omega t);$$  (14a)

the velocity potential is

$$\phi(x,z,t) = a\frac{\omega}{k}\exp(kz)\sin(kx - \omega t);$$  (14b)

the orbital velocities are

$$\tilde{u}_1 = u_x = a\omega\exp(kz)\cos(kx - \omega t) \quad \text{and} \quad \tilde{u}_3 = u_z = a\omega\exp(kz)\sin(kx - \omega t).$$  (14c)

Second, due to the orthogonality of the oscillating functions, from (14c) one may get

$$< \tilde{u}_1 \tilde{u}_3 > = 0.$$  (15)

Using the Prandtl approximation for the wave-induced turbulent motions in the form

$$\tilde{u}_{1,3}' = \lambda_{1,3}' \partial \tilde{u}_{1,3} / \partial x_3 \quad ,$$

for the same reason as in Eq. (15), one may find that

$$< \tilde{u}_1' \tilde{u}_3' > \approx 0.$$  (16)

Third, excluding the last (background) term in (13), and using Eqs. (15), (16), one gets the following wave-induced stress

$$< u_1 u_3 >_w = < \tilde{u}_1 u_3' > + < u_1' \tilde{u}_3 > \quad .$$  (17)

It is the wave-induced stress, $< u_1 u_3 >_w$, which gives the addition to the background turbulent viscosity given by Eq. (11). For the first time, the analog of Eq. (17) was postulated and analyzed in Qiao et al. (2004) (see Sect. 1).

## 2.3 The closure statements

New approach for the closure of term (17) has the following steps.

1) Due to the spatial homogeneity of the wave-induced turbulence, one may put that the both summands in (17) has the same value/ Thus,

$$< u_1 u_3 >_w \cong 2 < \tilde{u}_1 u_3' > \quad .$$  (18)

2) Using the Prandtl approximation for the background turbulence fluctuation, one has

$$u_3' = \lambda_3' \partial \bar{U} / \partial x_3,$$  (19)

where $\lambda_3'$ is the unknown, stochastic, background mixing-length, which has no relation to the wave motions, as far as $u_3'$ describes the background turbulence. In such a case, the main expression for the farther analysis is as

$$< u_1 u_3 >_w \cong 2 < \tilde{u}_1 \lambda_3' > \partial \bar{U} / \partial x_3.$$  (20)

3) By analogy with Eq. (11), from (20) it follows the implicit expression for the wave-induced turbulent viscosity, $B_v$ (in the notations by Qiao et al., 2004), of the form:

$$B_v \approx 2 < \tilde{u}_1 \lambda_3' > .$$  (21)

It is still remain to find the closure for statistically averaged value $< \tilde{u}_1 \lambda_3' >$. Note that Eq. (21) for $B_v$ differs from Eq. (2a) derived in Qiao et al. (2004). In our case, the background-turbulence mixing length, $\lambda_3^{'}$, is used, which cannot be expressed via any part of wave motions, as we have already mentioned from the very beginning of our consideration (Sect. 2.1).

4) Here it should be noted that function $B_v \approx 2 < \tilde{u}_1^{'} \lambda_3^{'} >$ has the following features:

(a) $B_v$ is the statistical moment, the final value of which can be postulated under some assumption (as it is used in the theory of turbulence) (Monin & Yaglom, 1971);

(b) $B_v$ is the linear function in the wave-induced turbulent fluctuation, $\tilde{u}_1^{'}$. From the physical point of view, one may expect that $B_v$ (at depth $z$) should also be the linear function in the local wave amplitude, $a(z)$, depending on $z$.

5) To go on, we propose to estimate statistical moment $< \tilde{u}_1^{'} \lambda_3^{'} >$, basing on the recent theory for the wind-induced drift current, derived in Polnikov (2018). According to his theory (Polnikov, 2010, 2018) and the empirical observations by Longo et al. (2012), the wavy air-sea interface has the three-layer structure: air-boundary layer (ABL), wave zone (WZ), and water-boundary layer (WBL) (or sub-surface boundary layer, in the notations by Longo et al. (2012) in Fig. 1).

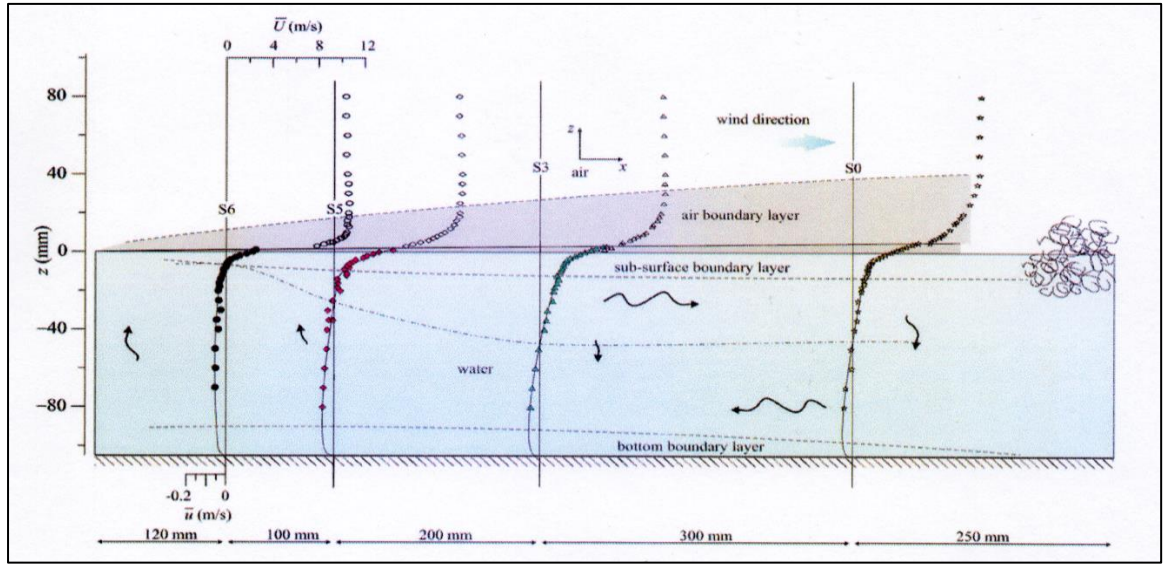

**Figure 1**. General scheme for the mean-flow distribution in the interface system (from Longo et al, 2012). The wave-zone is evidently shown between ABL and sub-surface BL

The main feature of this structure of the air-sea interface consists in that the vertical profile of the mean current, $\bar{U}(z)$, is linear in $z$ (Fig. 1). This fact is directly fixed in (Longo et al,, 2012). Due to this, the WZ has a meaning of the viscous layer located between the ABL and WBL.

Thus, from the point of view of the statistical hydrodynamics, in this zone the following conditions should take place:

(a) The wind-induced momentum flux, $\tau_w$, directed downward, is constant.

(b) The viscosity coefficient, $K_{tw}$, supporting the linear current profile, $\bar{U}(z)$, is constant.

Accepting the balance between $\tau_w$ and the vertical gradient of the mean current, $\partial \bar{U}_1 / \partial x_3$, typically used the turbulence theory (Monin & Yaglom, 1971),

$$\tau_w = K_{tw} \partial \overline{U}(z) / \partial z,$$

Polnikov (2018) has found that, in the WZ, the viscosity coefficient is

$$K_{tw} \approx c_{tw} u_* a_{|z=0}, \qquad (22)$$

where $c_{tw}$ is the dimensionless coefficient of the order of $10^{-2}$, and $u_*$ is the friction velocity in the ABL (for details, see original). Equation (22) means that the wave-induced turbulent viscosity in the WZ, $K_{tw}$, is the linear function in wave amplitude $a$, similar to the dependence expected for $B_v$.

The mentioned fact allows us to propose that the sought turbulent viscosity in the WBL, $B_v$, is the natural extension of the turbulent viscosity coefficient, $K_{tw}$, taking place in the WZ. Basing on this phenomenological assumption, one may state that

$$B_v(z) \approx K_{tw}(z), \qquad (23)\_$$

where $K_{tw}(z)$ is the analytical continuation of function $K_{tw}$ at the mean water level, $z = 0$, given by Eq. (22).

Thus, the explicit, depth-dependent expression for $B_v(z)$, in the case of monochromatic wave on the surface, obtains the form

$$B_v(z) \approx K_w(z) \approx c_v \cdot u_* a_0 \exp(kz), \qquad (24)$$

where $c_v$ is the dimensionless fitting coefficient of the order of $10^{-2}$ and $a_0$ is the wave amplitude at mean water level. The generalization of Eq. (24) for a spectrum of surface waves is given by the equation

$$B_v(z) = c_{Bv} \cdot u_* \left( \int S(\mathbf{k}) \exp(2kz) d\mathbf{k} \right)^{1/2}, \qquad (25)$$

where $c_{Bv}$ the fitting coefficient of the order of $10^{-2}$, similar to $c_v$ in Eq. (24).

### 2.4. Final remarks

Equations (24, 25) finalize the main derivations of the sought wave-induced vertical mixing function, $B_v$. As seen, the principal difference between Eq. (25) and result (3) by Qiao et al. (2004) is the much weaker power-dependence of $B_v$ on the local wave amplitude in Eq. (25). This result leads to the reasonably enhanced intensity for the wave-induced vertical mixing at depth $z$, due to the radical difference between exponents in Eqs, (3) and (25).

The predicted dependences, $B_v(a_0)$ and $B_v(u_*)$, can be verified in future by means of numerical simulations alike (Babanin & Chalikov, 2012; Skote & Henningson, 2002), and laboratory experiments alike (Dai et al., 2010). A successful verification of these dependences would justify the preference of new version of the model for the wave-induced mixing processes, with respect to the known one.

Some ideas of analytical verification dependences $B_v(a_0)$ and $B_v(u_*)$ are discussed below.

## 3 Discussion

Let us check the correspondence of result (25) to some known empirical dependences. As empirical data of a direct measuring viscosity coefficient $B_v$ (if any) are not known to us, we need to choose proper physical values for the experimental checking dependences $B_v(a_0)$ and $B_v(u_*)$. The most convenient of them is the rate of dissipation of the turbulent kinetic energy, $\varepsilon$, which is routinely measured in experiments (Anis & Moum, 1995; Gemmrich & Farmer, 2004, Jones & Monismith, 2009; Babanin & Hous, 2009, among others). For our aim, we may use the relation between the wave-induced part of the rate of dissipation of the turbulent kinetic energy, $\varepsilon_w$, and turbulent viscosity $B_v$, written in the form (Yuan et al., 2013)

$$\varepsilon_w \simeq B_v \left( \partial \tilde{u}_1 / \partial x_3 \right)^2 \quad .$$  (26)

Here, the vertical gradient of the wave-orbital velocity is used, as the approximation for qualitative estimations. Equation (26) allows an analytical checking the correspondence of Eqs. (24, 25) to the known dependences of $\varepsilon_w$ on wave amplitude $a_0$ and friction velocity $u_*$.

According to the measurements (e.g., Babanin & Hous, 2009), the following dependence $\varepsilon_w(a_0)$ takes place

$$\varepsilon_w \propto a_0^3 \quad .$$  (27)

Correspondence between ratios (25) and (27) becomes evident, if one takes into account that wave velocity $\tilde{u}_1$ in (26) is linear in wave amplitude $a_0$ (see Eq. 14c). Thus, the analytical checking dependence $B_v(a_0)$ is successful.

The known dependence of $\varepsilon_w$ on $u_*$ in WBL has the kind (e.g., Jones & Monismith, 2008)

$$\varepsilon_w \propto u_*^3 \quad .$$  (28)

From (25) and (14c), Eq. (26) results in the ratio (at the mean surface level)

$$\varepsilon_w \propto u_* a ( \omega_p k_p a )^2 \propto u_* \omega_p^6 a^3 \quad ,$$  (29)

where $\omega_p$ is the peak frequency of a wave spectrum, and the zero-subindex at wave amplitude $a$ is omitted for simplicity. To extract dependence $\varepsilon(u_*)$ from Eq. (29), one may use the well-known wave-growth dependences for the dimensionless wave energy and peak frequency on the dimensionless fetch, in the fetch-limited case. They are as follows (Komen et al, 1994):

$$\frac{a^2 g}{u_*^4} \propto \frac{Xg}{u_*^2} \qquad \text{and} \qquad \frac{\omega_p u_*}{g} \propto \left( \frac{Xg}{u_*^2} \right)^{-1/3} \quad ,$$  (30)

where $X$ is the dimensional fetch, and $g$ is the gravity acceleration. From the first Eq. (30), one gets that wave amplitude has the following dependence: $a(u_*) \propto u_*$. From the second Eq. (30), it follows that $\omega_p(u_*) \propto (u_*)^{-1/3}$. Finally, Eq. (29) gets the form

$$\varepsilon_w \propto u_*\omega_p^6 a^3 \propto u_*^2 \quad ,$$  (31)

what corresponds reasonably to empirical dependence (28), taking into account inevitable empirical errors of estimates for powers in Eqs. (27), (28), (30).

Thus, the encouraging results of analytical checking the correspondence of Eqs. (24, 25) to known empirical dependences (27) and (28) open a way to its experimental verification. The latter could be realized by the estimating empirical dependences $B_v(a_0)$ and $B_v(u_*)$ in the tank experiments similar to ones describes in (Babanin & Hous, 2009; Dai et al., 2010).

In addition to the said, we would like to mention the following. In our mind, experimental verification dependences $B_v(a_0)$ and $B_v(u_*)$ can be executed by two ways. The first, indirect way could be based on a measuring the rate of dissipation of the turbulent kinetic energy, $\varepsilon_w$, with the next using the r.h.s of Eq. (29) for the comparison, as presented above. The second, direct way could be realized by means of estimating the wave-induced vertical mixing function $B_v$ via a measuring the rate of a spatial spreading of a small color ink-drop in the water layer below the deepest wave troughs. To this aim, one can use the Einstein formula, keeping in mind the physical similarity between $B_v$ and the diffusion coefficient for passive particles in the water (Monin & Yaglom, 1971). The proper technique needs its own specification, though, it could be easily elaborated if needed.

## 4 Conclusions

The model for the vertical turbulent-mixing function is derived, which predicts the enhanced impact of the wind-wave motions on the mixing in the upper ocean. It means that surface wind waves can make a much more impact on all geophysical processes related to a vertical mixing in oceans than it is known already (e.g., Qiao et al., 2016; Aijaz et al., 2017; Walsh et al., 2017).

The model is based on the following three grounds.

First, in the Navier-Stokes equations, the total current is decomposed into the four constituents, including the mean current, wave-orbital motions, wave-induced-turbulent and background-turbulent currents (Eq. 5). This allows to separate the wave-induced Reynolds stress, $R_w$, from the background one, $R_b$ (Eq. 9).

Second, to close wave-induced stress $R_w$, the Prandtl approximation for the background turbulence fluctuation is used, resulting in the implicit expression for the wave-induced vertical mixing coefficient, $B_v$ , valid in the upper layer (Eq. 20).

Third, the expression for $B_v$ is specified, basing on the author's results for the turbulent viscosity, $K_{tw}$, taking place in the wave zone, located between the air- and water-boundary layers of the air-sea interface (item 5) in Sect. 2.3). The sought turbulent viscosity in the upper water layer. $B_v$. is proposed to be the analytical continuation of the function describing $K_{tw}$. Eventually, $B_v$ is found as the linear function in both wave amplitude $a(z)$ and friction velocity $u_*$ in the air (Eqs. 24, 25).

The found result for $B_v(a)$ means the possibility of the enhanced impact of waves on the vertical mixing, compared with the known cubic dependence of $B_v(a)$ described by Eq. (3) (Qiao et al., 2004). The enhancing is stipulated by the exponential decay of the amplitude for the wave-orbital motion: $a(z) \sim exp(kz)$.

The analytically predicted dependences $B_v(a)$ and $B_v(u_*)$ can be verified empirically by estimating them in the tank-experiments alike (Babanin & Hous, 2009; Dai et al, 2010). They also could be estimated numerically by means of the direct numerical simulations alike (Skote & Henningson, 2002), for a further comparison with the presented analytical derivations (Eq. 29).

One may hope that the mentioned forefront researches in this direction open a way for numerous geophysical applications of the result found here, similarly to ones have followed the pioneering paper by Qiao et al. (2004).

**Acknowledgments** The author is grateful to Profs. N. Huang and D. Dai for the useful advises given while a preliminary discussion of the problem under consideration. The work was supported by the Russian Foundation for Basic Research, grant No. 18-05-00161.